\documentclass[fleqn,usenatbib]{mnras}

\usepackage{newtxtext,newtxmath}
\usepackage[T1]{fontenc}
\usepackage{natbib}
\DeclareRobustCommand{\VAN}[3]{#2}
\let\VANthebibliography\thebibliography
\def\thebibliography{\DeclareRobustCommand{\VAN}[3]{##3}\VANthebibliography}

\usepackage{graphicx}	% Including figure files
\usepackage{amsmath}	% Advanced maths commands
\usepackage{newtxtext,newtxmath}
\usepackage[T1]{fontenc}
\usepackage{ae,aecompl}
\usepackage{epsfig}
\usepackage{amsmath, amsfonts, epsfig, xspace}
\usepackage{natbib}
\usepackage{deluxetable}
\usepackage{rotating}
\usepackage{graphicx}
\usepackage{longtable}
\usepackage{multicol}
\usepackage{booktabs}
\usepackage{dcolumn}
\usepackage{longtable}
\usepackage{changepage}
\newcolumntype{d}[1]{D{.}{.}{#1}}
\usepackage{float}
\title[INOV of 6 TeV blazars]{Intranight optical variability of TeV blazars with parsec-scale jets dominated by slow-moving radio knots}

\author[Negi et al.]{
Vibhore Negi$^{1,2}$,\thanks{E-mail:vibhore.negi18@gmail.com}
Gopal-Krishna$^{3}$,
Hum Chand$^{4}$,
and Silke Britzen$^{5}$
\\\\
$^{1}$Aryabhatta Research Institute of Observational Sciences (ARIES), Manora Peak, Nainital, 263002, India\\
$^{2}$Department of Physics, Deen Dayal Upadhyaya Gorakhpur University, Gorakhpur, 273009, India\\
$^{3}$UM-DAE Centre for Excellence in Basic Sciences, Vidyanagari, Mumbai, 400098, India\\
$^{4}$Department of Physics and Astronomical Sciences, Central University of Himachal Pradesh (CUHP), Dharamshala, 176215, India\\
$^{5}$Max-Planck-Institut f.Radioastronomie, Auf den Huegel 69, 53121 Bonn, Germany\\
}

\date{Accepted XXX. Received YYY; in original form ZZZ}

\pubyear{2023}

\begin{document}
\label{firstpage}
\pagerange{\pageref{firstpage}--\pageref{lastpage}}
\maketitle

% Abstract of the paper
\begin{abstract}
BL Lac objects detected at TeV energies preferentially belong to the subclass called ‘high-frequency-peaked’ BL Lacs (HBLs). Parsec-scale radio jets in these TeV-HBLs often show dominant, slow moving radio knots that are at most mildly superluminal. We report the first systematic campaign to characterise the Intra-Night Optical Variability (INOV) of TeV-HBLs using a representative sample of 6 such sources, all showing a fairly high degree of optical polarization. Our campaign consists of high-sensitivity monitoring of this sample in 24 sessions of more than 3 hour duration each. For these TeV-HBLs, we find a striking lack of INOV and based on this, we discuss the importance of superluminal motion of the radio knots vis-a-vis the optical polarization, as the key diagnostic for INOV detection.
\end{abstract}
%%%%%%%%%%%%%%%%%%%%%%%%%

%%%%%%%%%%%%%%%%%%%%%%%%%%%%%%%%%%%%%%%%%%%%%%%%%%

%%%%%%%%%%%%%%%%% BODY OF PAPER %%%%%%%%%%%%%%%%%%

\section{Introduction}
\label{sec:introduction}
Quasars whose observed radiation at centimetre and shorter wavelengths
arises predominantly from a jet producing nonthermal radiation relativistically beamed towards the observer, are termed as
blazars. They exhibit flux variability across the electromagnetic spectrum on diverse time scales \cite[reviewed, e.g., by ][]{1995ARA&A..33..163W,1997ARA&A..35..445U,2016Galax...4...37M, 2019ARA&A..57..467B}. Spectroscopically, blazar population is subdivided between `broad-line emitting’ flat-spectrum radio quasars (FSRQs) and BL Lac objects (BL Lacs) showing an almost featureless optical/UV spectrum \citep{1991ApJ...374..431S}, excepting the few cases for which spectral features due to host galaxy have been detected. Blazars with synchrotron emission peaking 
at high frequencies, between UV and X-ray bands, i.e., $\nu_{syn}^{peak} >10^{15}$ Hz are most commonly BL Lacs and these are called HBLs (e.g., reviews by \citealt{1993ARA&A..31..473A, 1995MNRAS.277.1477P, 1995PASP..107..803U, 2016A&ARv..24...10T}). Compared to the BL Lacs with synchrotron spectra peaking below $\sim 10^{14}$ Hz (called LBLs, \citealt{1995MNRAS.277.1477P}; \citealt{2010ApJ...716...30A}), HBLs are preferentially detected at TeV energies and a few dozen such TeV-HBLs  have been catalogued \citep{2008ICRC....3.1341W}. HBLs typically have modest intrinsic radio luminosities, as compared to LBLs and are usually hosted by `low-excitation radio galaxies’ (LERGs), whose central engines are powered by radiatively inefficient gas accretion on to the central supermassive black holes (SMBH, see, e.g.; \citealt{2001MNRAS.327..739G,2005A&A...432..401G,2009MNRAS.396L.105G,2011ApJ...740...98M,2012MNRAS.420.2899G,2014MNRAS.445...81S}).
%; also \citealt{1989ApJ...340..181G, 1996ApJ...463..444S}). {\bf CAN WE LIMIT TO 4 PAPERS?}
It is commonly believed that the parent
(i.e., misaligned) population of BL Lacs is Fanaroff-Riley type I (FR I, \citealt{1974MNRAS.167P..31F}) radio galaxies \citep[e.g.,][]{1984ApJ...279...93W,1989LNP...334..401B,1989ApJ...336..606B,1993ARA&A..31..473A,1995PASP..107..803U}. \citet{1999A&A...349...77C} showed that relativistic beaming, rather than obscuration, of the nuclear jets can account for the $10 - 10^{4}$ difference in radio and optical luminosities between BL Lacs and FR I radio  galaxies and the required beaming typically needs bulk Lorentz factors of just a few \citep[e.g.,][]{1999A&A...349...77C, 1999MNRAS.306..513L, 1991ApJ...371...60U, 2003A&A...403..889T, 2003MNRAS.338..176H}. 
The transverse dual-velocity structure of jets was independently hypothesized by \citet{2000A&A...358..104C} taking into account the observed correlation between the radio and the optical core luminosity in FR I radio galaxies and BL Lacs. A more direct evidence for  the `spine-sheath’ jet scenario comes from the observed limb-brightening of parsec-scale jets in several lower-luminosity radio sources, e.g., the HBLs Mrk 421 and Mrk 501 (\citealt{ 2009ApJ...690L..31P, 2010ApJ...723.1150P, 
2004ApJ...600..127G, 2006ApJ...646..801G, 2021MNRAS.503.3145B}, %{\color{blue} Britzen et al., 2023 (in press)}, 
\citealt{2023Universe...9L..00B, 2021NatAs...5.1017J, 2021A&A...654A..27B}),
%2016A&A...585A..33B, 2016A&A...587A..52M, 2018A&A...616A.188K
% {\color{red}(Hervet et al. 2018)}; 
and also in some kiloparsec-scale jets \citep{1989ApJ...340..698O, 1998ApJ...507L..29S, 2011MNRAS.417.2789L}. %Plausibly, the limb-brightening is manifesting a transverse stratification of bulk velocity and/or magnetic field in the jet. 

Another well-documented manifestation of blazar activity is their Intra-Night Optical Variability (INOV) (\citealt{2018BSRSL..87..281G} and references therein; \citealt{2021MNRAS.507L..46M, 2023MNRAS.518L..13G}). At least in the context of blazars, INOV is believed to arise mainly due to a combination of two factors: (i) generation of turbulence within the jet plasma whose synchrotron emissivity and fractional polarization can increase while passing through one or more shocks (\citealt{2014ApJ...780...87M} and references therein; \citealt{2015JApA...36..255C}; also, \citealt{2012A&A...544A..37G,1980MNRAS.193..439L}) and (ii) Doppler factor $\delta_{j}$ of the post-shock turbulent jet plasma.
% (the flow may pass through additional shocks downstream). 
While the former condition is crucial for inducing micro-variability, the latter can play a key role in making it detectable (via a Doppler boost). A strong dependence of INOV on fractional optical polarization ($p_{opt}$) was first established by \citet{2012A&A...544A..37G}, who showed that a flat/inverted radio spectrum by itself does not ensure a strong tendency for INOV.
The question remains whether $p_{opt}$ alone suffices, or the other factor mentioned above, namely, a strong beaming as inferred from the apparent speed of the VLBI radio knots, also plays a dominant role? Since, to our knowledge, no systematic investigation of this issue has been reported, we have carried out an INOV campaign targeting representative sample of 6 TeV blazars. The crucial aspect of these blazars is that even though they fall within the high polarization class (HPQ), their nuclear jets are dominated by radio knots showing at most mildly superluminal motion which
is statistically consistent with zero radial velocity from the core, in a majority of cases (see, Table~\ref{tab:sample_table}). The selection of the sample representing this extreme subset of HPQs is described in Section~\ref{sec:Sample}. The observations and data reduction procedures are outlined in Sections~\ref{sec:photomteric_monitoring} \& \ref{sec:analysis}. Section~\ref{sec:results} presents the results together with a brief discussion. Our main conclusions are summarised in Section~\ref{sec:conclusions}. \par
%%%%%%%%%%%%%%%%%%%%%%%%%%%%%%%%%
%%%%%%%%%%%%%%%%%%%%%%%%%%%%%%%%
%%%%%%%%%%%%%%%%%%%%%%%%%%%%%%%%
\section{Sample selection}
\label{sec:Sample}
For the purpose of (optical) differential aperture-photometry, the present sample of 6 TeV-HBLs (Table~\ref{tab:sample_table}) has been drawn from the VLBI data published in \citet{2018ApJ...853...68P} for a sample of 38 TeV-HBLs. We imposed a limit of $z \gtrsim 0.3$, in order to minimise the relative contribution from the host galaxy and thereby the possibility of claiming spurious INOV detection, in case the `point spread function' (PSF) changes during the monitoring session \citep{2000AJ....119.1534C}. 
%The latter can lead to spurious INOV due to gradients in `seeing’ during the AGN monitoring session. This is because a changing `point spread function’ (PSF) leads to variation in the relative fraction of light from the point-like blazar and its (extended) host galaxy to be included inside the aperture \citep{1991AJ....101.1196C, 2000AJ....119.1534C}. {\bf 
This resulted in exclusion of 30 of the sources. 
%\item Out of the 38 sources, 30 with $z \leq 0.3$ were excluded, in order to minimise the contribution of the host galaxy which can vary with the changing seeing disk, leading to the possibility of spurious claims of INOV detection (\citealt{2000AJ....119.1534C}). 
Another two sources got discarded due to the second filter, imposed by observational considerations, namely (i) declination > 0 and (ii) 
%For reasons of observational accessibility,  we rejected 1 more source south of declination ($\delta \leq0$).
%\item To the remaining 7 sources, we applied a cut in $r-$band magnitude 
$m_{r} \leq 17.50-$mag, taking $m_{r}$ from the Pan-STARRS DR1 (\citealt{2016arXiv161205560C}). 
%This was from the consideration of sensitivity attainable in monitoring with the 1-metre class telescope (DFOT, sect. 3) of ARIES used by us. \\
This left us with a sample of 6 VLBI monitored TeV-HBLs (Table~\ref{tab:sample_table}). It is seen that for only two of the 6 sources, J0507+6737 and J1427+2348, the estimated $\beta_{app}$ deviates from zero by more than $\sim$ 2$\sigma$, the most deviant being J0507+6737 for which the deviation is significant at 5.1$\sigma$ (but, even in this case, the motion is only mildly superluminal).

\begin{table*}%[!h]
  \centering
%\begin{minipage}{110mm}
    \caption{The sample of 6 TeV-HBLs selected from \citet{2018ApJ...853...68P}.}
    \resizebox{\textwidth}{!}{%
%\begin{tabular}{cccccccccccc}
\begin{tabular}{p{2.25cm}p{1.5cm}p{1.5cm}p{1cm}p{1cm}p{2.5cm}p{5cm}p{3cm}p{1.0cm}p{0.05cm}p{0.001cm}}
  %\scriptsize
  \hline\\
  %Blazar  &RA (J2000) & Dec (J2000) & Redshift & $m_{r}^\dagger$    & Apparent$^{\ast}$    & Minimum$^{\ddagger}$  &   Maximum$^{\ddagger}$   &   Mean$^{\ddagger}$   & Median$^{\ddagger}$  &  Number of$^{\ddagger}$  \\
  %Name &    &   &   &   & Jet Speed  & polarization (\%)  & polarization (\%) & polarization (\%)  & polarization (\%) &    polarization  &  \\
  % & hh:mm:ss   &$^\circ$: $^\prime$: $^{\prime\prime}$   & $z$ (NED)	&   &$\beta_{app}$ & (JD of observation)  & (JD of observation)  & &   & Measurements \\
  \multicolumn{1}{c}{Blazar}  &\multicolumn{1}{c}{RA (J2000)} & \multicolumn{1}{c}{Dec (J2000)} & \multicolumn{1}{c}{Redshift$^{\triangle}$} & \multicolumn{1}{c} {$m_{r}^\dagger$}    & \multicolumn{1}{c} {Apparent$^{\ast}$}    & \multicolumn{1}{c} {Minimum$^{\ddagger}$}  &    \multicolumn{1}{c} {Maximum$^{\ddagger}$}   &    \multicolumn{1}{c} {Mean$^{\ddagger}$ }  & \multicolumn{1}{c}{Median$^{\ddagger}$}  &   \multicolumn{1}{c} {Number of$^{\ddagger}$}  \\
  \multicolumn{1}{c}{Name} &    &   &   &   & \multicolumn{1}{c}{Jet Speed}  & \multicolumn{1}{c}{polarization (\%)}  & \multicolumn{1}{c}{polarization (\%)} & \multicolumn{1}{c}{polarization (\%)}  & \multicolumn{1}{c}{polarization (\%)} &    \multicolumn{1}{c}{polarization} \\
 	%& \multicolumn{1}{c}{hh:mm:ss}   & \multicolumn{1}{c}{$^\circ$: $^\prime$: $^{\prime\prime}$}   & \multicolumn{1}{c}{$z$ (NED)}	&   & \multicolumn{1}{c}{$\beta_{app}$} & \multicolumn{1}{c}{(JD of observation)}  & \multicolumn{1}{c}{(JD of observation)}  & &   & \multicolumn{1}{c}{Measurements} \\
 	& \multicolumn{1}{c}{hh:mm:ss}   & \multicolumn{1}{c}{$^\circ$: $^\prime$: $^{\prime\prime}$}   & \multicolumn{1}{c}{$z$}	&   & \multicolumn{1}{c}{$\beta_{app}$} & \multicolumn{1}{c}{(JD of observation)}  & \multicolumn{1}{c}{(JD of observation)}  & &   & \multicolumn{1}{c}{Measurements} \\
 \hline\\

\multicolumn{1}{c}{J0136+3905}	&  \multicolumn{1}{c}{01:36:32.60} & \multicolumn{1}{c}{+39:05:59.2} &  \multicolumn{1}{c}{0.750$^{\textbf a}$} & \multicolumn{1}{c}{15.83}  & \multicolumn{1}{c}{0.98$\pm$1.13$^{1}$ (< 1$\sigma$)}   &   \multicolumn{1}{c}{1.4$\pm$0.2}	&  \multicolumn{1}{c}{4.1$\pm$0.5} & \multicolumn{1}{c}{2.37$\pm$0.14}  &   \multicolumn{1}{c}{1.9} &  \multicolumn{1}{c}{7} \\
\multicolumn{1}{c}{(RGB J0136+391)} &   &   &   &   & \multicolumn{1}{c}{1.82$\pm$3.05$^{1}$ (< 1$\sigma$)}  &   \multicolumn{1}{c}{(2457336.4624)}	&  \multicolumn{1}{c}{(2456853.5927)}    & &  &   \\\\

\multicolumn{1}{c}{J0416+0105}		&  \multicolumn{1}{c}{04:16:52.49} & \multicolumn{1}{c}{+01:05:23.9} &  \multicolumn{1}{c}{0.287$^{\textbf b}$} & \multicolumn{1}{c}{16.56}  & \multicolumn{1}{c}{0.03$\pm$0.58$^{1}$(< 1$\sigma$)}   &   \multicolumn{1}{c}{4.8$\pm$0.9}	&  \multicolumn{1}{c}{8.5$\pm$0.4} & \multicolumn{1}{c}{6.28$\pm$0.28}  &   \multicolumn{1}{c}{5.9} &  \multicolumn{1}{c}{4} \\
\multicolumn{1}{c}{(1ES 0414+009)}  &   &   &   &   &   &   \multicolumn{1}{c}{(2456941.5946)}	&  \multicolumn{1}{c}{(2457264.6115)}    & &  &   \\\\

\multicolumn{1}{c}{J0507+6737}		&  \multicolumn{1}{c}{05:07:56.15} & \multicolumn{1}{c}{+67:37:24.3} &  \multicolumn{1}{c}{0.314$^{\textbf c}$} & \multicolumn{1}{c}{17.01}  & \multicolumn{1}{c}{2.23$\pm$0.44$^{1}$(5.1$\sigma$)}   &   \multicolumn{1}{c}{2.5$\pm$0.6}	&  \multicolumn{1}{c}{6.5$\pm$0.4} & \multicolumn{1}{c}{4.15$\pm$0.14}  &   \multicolumn{1}{c}{4.0} &  \multicolumn{1}{c}{15} \\
\multicolumn{1}{c}{(1ES 0502+675)}  &   &   &   &   & \multicolumn{1}{c}{7.92$\pm$3.51$^{1}$ (2.3$\sigma$)}  &   \multicolumn{1}{c}{(2456592.5206)}	&  \multicolumn{1}{c}{(2456937.6025)}    & &  &   \\\\

\multicolumn{1}{c}{J0650+2502}	&  \multicolumn{1}{c}{06:50:46.49} & \multicolumn{1}{c}{+25:02:59.6} &  \multicolumn{1}{c}{0.410$^{\textbf d}$} & \multicolumn{1}{c}{14.61}    & \multicolumn{1}{c}{-3.13$\pm$1.67$^{1}$( 1.9$\sigma$)}   &   \multicolumn{1}{c}{6.3$\pm$0.4}	&  \multicolumn{1}{c}{9.4$\pm$0.4} & \multicolumn{1}{c}{8.70$\pm$0.20}  &   \multicolumn{1}{c}{9.3} &  \multicolumn{1}{c}{5} \\
\multicolumn{1}{c}{(1ES 0647+250)}  &   &   &   &   &   &   \multicolumn{1}{c}{(2456980.4321)}	&  \multicolumn{1}{c}{(2457337.5394)}    & &  &   \\\\

\multicolumn{1}{c}{J1427+2348}	&  \multicolumn{1}{c}{14:27:00.39} & \multicolumn{1}{c}{+23:48:00.0} &  \multicolumn{1}{c}{ 0.604$^{\textbf e}$} & \multicolumn{1}{c}{14.11}   &  \multicolumn{1}{c}{2.83$\pm$0.89$^{2}$( 3.2$\sigma$)}   &   \multicolumn{1}{c}{0.3$\pm$0.3}	&  \multicolumn{1}{c}{10.6$\pm$0.3} & \multicolumn{1}{c}{5.76$\pm$0.09}  &   \multicolumn{1}{c}{5.9} &  \multicolumn{1}{c}{12} \\
\multicolumn{1}{c}{(PKS 1424+240)} &   &   &   &   &   &   \multicolumn{1}{c}{(2456780.4278)}	&  \multicolumn{1}{c}{(2457228.2769)}    & &  &   \\\\

\multicolumn{1}{c}{J1555+1111} 	&  \multicolumn{1}{c}{15:55:43.04} & \multicolumn{1}{c}{+11:11:24.4} &  \multicolumn{1}{c}{ 0.360$^{\textbf f}$} & \multicolumn{1}{c}{14.28}   & \multicolumn{1}{c}{0.95$\pm$1.09$^{3}$(< 1$\sigma$)}   &   \multicolumn{1}{c}{0.2$\pm$0.3}	&  \multicolumn{1}{c}{9.6$\pm$0.3} & \multicolumn{1}{c}{3.29$\pm$0.02}  &   \multicolumn{1}{c}{2.9} &  \multicolumn{1}{c}{284}  \\
\multicolumn{1}{c}{(PG 1553+113)}   &   &   &   &   &   &   \multicolumn{1}{c}{(2456519.2927)}	&  \multicolumn{1}{c}{(2456914.227)}    & &  &   \\\\

 \hline
 \multicolumn{11}{l}{ {$^{\triangle}$ References for the redshift: (a). \citealt{2015A&A...575A..21N}; (b). \citealt{2019A&A...632A..77C}; (c). \citealt{2021AJ....161...45O}; (d). \citealt{2011A&A...534L...2K}; (e). \citealt{2017ApJ...837..144P}; (f). \citealt{2016ApJ...818..113N}. The lower limits on redshifts for}  }\\
 \multicolumn{11}{l}{ ~~~  J0136+3905, J0650+2502, J1427+2348, and J1555+1111 are also available in the ZBLLAC database based on high SNR optical spectra \citep{2020ApJS..250...37L}, consistent with their aforementioned  redshifts.}\\
  %\multicolumn{11}{l}{ {\bf ~~~ on the detection of intervening absorption systems.} }\\
 \multicolumn{11}{l}{ {\bf $^{\dagger}$} $ m_{r}$ is taken from Pan-STARRS survey \citep{2016arXiv161205560C}.}\\
 \multicolumn{11}{l}{ {\bf $^{\ast}$ } References for the jet speed: 1. \citet{2018ApJ...853...68P}; 2. \citet{2019ApJ...874...43L}; 3. \citet{2012arXiv1205.2399T}. The speeds are given in units of c and have been derived from the observed proper motions of the dominant VLBI knots.}\\
\multicolumn{11}{l}{ {\bf $^{\ddagger}$ } The polarization data has been taken from \citet{2021MNRAS.501.3715B}. The ‘mean’ values are our estimates
based on the RoboPol data \citep{2021MNRAS.501.3715B}.}\\\\
\hline
  \end{tabular}
 }
 \label{tab:sample_table}
 %\end{minipage}
\end{table*}
%******************************************

%%%%%%%%%%%%%%%%%%%%%%%%%%%%%
%%%%%%%%%%%%%%%%%%%%%%%%%%%%%
\section{The Monitoring and data reduction}
\label{sec:photomteric_monitoring}
 The sample of 6 TeV-HBLs was monitored in Johnson-Cousins R-band in 24 sessions (i.e., 4 sessions per source), using the 1.3-metre Devasthal Fast Optical Telescope (DFOT; \citealt{2011CSci..101.1020S}) located at Devasthal station of ARIES (India). The images were recorded on a Peltier-cooled ANDOR CCD having 2k $\times$ 2k ( 0.53 arcsec pixel$^{-1}$) pixels, covering a field of view of 18.5 $\times$ 18.5 arcmin$^{2}$. The CCD detector has a gain of 2 e$^{-}$ per analog-to-digital unit (ADU) and a readout noise of 7e$^{-}$ at a speed of 1000 kHz. In each session, one target blazar was monitored continuously for minimum 3 hours, with a typical exposure of 1.5–5 min per frame. 
 %The typical seeing value during our monitoring sessions was $\sim$ 2-6 arcsec (online Figs 1, 2, and 3).

%\subsection{Data Reduction}
%\label{sec:data_reduction}
The pre-processing and cleaning of the CCD frames was done following the standard procedures in IRAF. The instrumental magnitude of the blazar and the two (steady appearing) comparison stars contained in all the CCD frames taken in the session were determined by aperture photometry \cite[see][]{1987PASP...99..191S,1992ASPC...25..297S}, using the DAOPHOT II (Dominion Astronomical Observatory Photometry II) package. The PSF was estimated by averaging the full width at half-maximum (FWHM) of the Gaussians fitted to the brightness profiles of five moderately bright stars within each frame, and aperture radius was set equal to two times the PSF (see e.g., \citealt{2019MNRAS.483.3036O}). The variation of PSF during each session is plotted in the bottom panel in the online Figures~S1-S6. For each session, we then derived differential light curves (DLCs) for all pairs involving the target blazar and the chosen two comparison stars (Figures~S1-S6 and Tables~S1 and S2 available online as Supporting Information).
%%%%%%%%%%%%%%%%%%%%%%%%%%%%%
%%%%%%%%%%%%%%%%%%%%%%%%%%%%%
 \section{Statistical Analysis}
 \label{sec:analysis}
 To ascertain the presence of INOV in our TeV-HBL sample, we applied the widely used $F_{\eta}$ test \cite[][]{2010AJ....139.1269D}, following the basic procedure described in \citet{2019MNRAS.489L..42M} and \citet{2022MNRAS.511L..13C}. The two steady comparison stars were chosen by inspecting several star-star DLCs derived for each session and the $F_{\eta}$ test was applied to the DLCs of the target blazar relative to the two comparison stars (whose basic parameters are listed in the online Table~S2). The $F$-values for the two blazar DLCs of a session are computed as:
 
 \begin{equation} 
\label{eq.ftest2}
F_{1}^{\eta} = \frac{Var(q-s1)}
{ \eta^2 \sum_{i=1}^{N}\sigma^2_{i,err}(q-s1)/N},  \\
\hspace{0.1cm} F_{2}^{\eta} = \frac{Var(q-s2)}
{ \eta^2 \sum_{i=1}^{N}\sigma^2_{i,err}(q-s2)/N}  \\
\end{equation}

%\vspace{-0.2in}
where $Var (q - s1)$ and $Var (q - s2)$ are the variances of the two DLCs of the target blazar,
% relative to the selected two comparison stars, 
and $\sigma_{i, err}(q - s1)$ \& $\sigma_{i,err}(q - s2)$ represent the rms error returned by DAOPHOT on the $i^{th}$ data point in a DLC of the target blazar.
%, relative to the two comparison stars. 
N is number of data points in the DLCs and the scaling factor ${\eta}$ = 1.54 \citep{ 2013JApA...34..273G,1995MNRAS.274..701G}. 
%It must be noted that several studies have reported the photometric errors returned by DAOPHOT to be underestimated by a factor ${\eta}$ \cite[][]{1995MNRAS.274..701G,2004JApA...25....1S,2005MNRAS.358..774B}  for which an accurate estimate of  $1.54 \pm 0.05$ was made by \citet{2013JApA...34..273G}, analysing a large set of 262 DLCs of pairs of steady comparison stars, monitored in 262 intranight sessions of minimum duration 3 h each, targeted at quasars/blazars.  
Online Table~S2 (Column 5) compares the computed values of $F_{\eta}$ for the two blazar DLCs of each session, with the critical value of {\it F} ($=F_{c}^{\alpha}$) estimated for that session. The values of $\alpha$ are set at 0.05 and 0.01, corresponding to 95 per cent and 99 per cent confidence levels for INOV detection. If the computed $F_{\eta}$ for a DLC of the target blazar exceeds $F_{c}^{\alpha}$, the null hypothesis (i.e., no variability) is discarded at the corresponding confidence level. Thus a DLC is classified as variable (`V') if the computed $F_{\eta} \geq F_{c}$ (0.99); probably variable (`PV') if the F${\eta}$ falls between 
$F_{c}$(0.95) and $F_{c}$(0.99); and non-variable (`NV') if $F_{\eta} \leq F_{c}$(0.95). Note that the target blazar in a session is designated as variable (V) only if both its DLCs (relative to the two comparison stars) belong to the `V' category, and `NV' if any of the two DLCs is of `NV' type. The remaining sessions are designated `PV'. The last column of the online Table~S2 lists the session's averaged photometric accuracy, the `Photometric Noise Parameter' (PNP) = {$\sqrt { \eta^2\langle \sigma^2_{i,err} \rangle}$ }, where ${\eta}$ = 1.54.

%Column 11 in table 3 lists the mean of the variability amplitudes ${\psi)$ of the target blazar found for its two DLCs in a session:

%\begin{equation}
%\psi= \sqrt{({A_{max}}-{A_{min}})^2-2\sigma^2}
%\end{equation}
%where $A_{max}$ and $A_{min}$ are the maximum and minimum values in the source-star DLC and $\sigma^2=\eta^2<\sigma^2_{q-s}>$, where, $\sigma^2_{q-s}$ is the mean square %rms error for the data points in the DLC. The error underestimation factor  ${\eta=1.54}$ (see above). \par
%\subsection{The duty cycle of variability}
%\label{sec:duty_cycle}

%The INOV duty cycle (DC) for our sample of 6 TeV blazars was computed as \citep{1999A&AS..135..477R}:
%\begin{equation} 
%DC = 100\frac{\sum_{i=1}^{N} K_i(1/\Delta t_i)}{\sum_{i=1}^{N}(1/\Delta t_i)} {\rm \%} 
%\label{eq:dc} 
%\end{equation}
%Here $\Delta t_i = \Delta t_{i,obs}(1+z)^{-1}$ is the intrinsic rest frame monitoring duration of the $i^{th}$ session, obtained by correcting for redshift, $z$ of the %source ($\Delta t_{i,obs}$ is listed in column 4 of the online Table 3). $K_i$ was taken as 1 for a positive detection of variability in $i^{th}$ session, otherwise, $K_i$ %= 0.\par
%%%%%%%%%%%%%%%%%%%%%%%%%%%%%
\begin{figure}
\includegraphics[width=0.45\textwidth,height=.35\textheight]{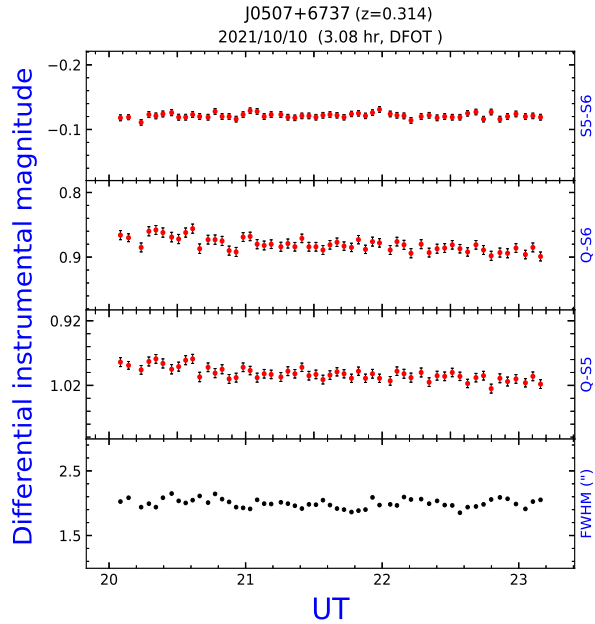}\\
  \vspace{-0.25in}
  \caption{Differential light curves (DLCs) for the TeV-HBL J0507+6737 on 2021-10-10, on which the blazar showed a hint of fading by $\sim$2.5\% over 3 hours, relative to both comparison stars. The upper plot gives the comparison star-star DLC and the middle two plots give the two blazar-star DLCs as defined in the labels on the right side. The lowest plot shows the seeing (PSF) variation during the session.}
\label{fig:all_dlc_J0507_PV} 
\end{figure}
%%%%%%%%%%%%%%%%%%%%%%%%%%%%%
%%%%%%%%%%%%%%%%%%%%%%%%%%%%%
%%%%%%%%%%%%%%%%%%%%%%%%%%%%%
\section{Results and Discussion}
\label{sec:results}
The present observations were mostly made under good sky conditions, using a 1.3-metre telescope located at a good site. With a typical threshold of $\psi \sim$ 2\% for INOV detection, these observations compare well, both in sensitivity and cadence, with practically all other INOV observations reported in the literature. Yet, strikingly, INOV was not detected in any of the 24 monitoring sessions targeting our sample of 6 TeV-HBLs (online Figures~S1-S6; Table~S2). One possible exception is the session on 2021-10-10, during which a hint of gradual fading by $\sim$ 2.5\% over 3 hours was noticed for the $z$ = 0.314 blazar J0507+6737. The fading was observed relative to both comparison stars which themselves remained steady throughout that session, as did the PSF (Fig.~\ref{fig:all_dlc_J0507_PV}). Interestingly, this is the only blazar in our sample for which \citet{2018ApJ...853...68P} have reported a (mildly) superluminal motion at a high confidence level ($\beta_{app}$  = 2.23 $\pm$0.44c, i.e., 5.1$\sigma$, see Table~\ref{tab:sample_table}). Even taking, conservatively, this possible INOV detection as confirmed (despite its formal classification being `non-variable', see online Table~S2), the INOV duty cycle for our sample would still be only $\sim$ 4\%. This is miniscule in comparison to (i) the INOV DC of  $\sim$60\% found for the sample of 13 TeV detected LBLs/FSRQs, which too mostly lie at $z$ > 0.3 \citep{2011MNRAS.416..101G}, and (ii) the INOV DC of $\sim$ 60 - 70\% generally found for LBLs \citep{1998A&A...329..853H, 2017ApJ...844...32P}. 
%{\color{red}(CHECK THIS CLAIM, and also check \citealt{2002A&A...390..431R, 2004JApA...25....1S, 2013MNRAS.435.1300G, 2021Galax...9..114W})}. 
Thus, TeV-HBLs with parsec-scale jets dominated by subluminal (or mildly superluminal) radio knots, appears to be an extreme subclass of blazars with an INOV duty cycle bordering on zero. While such a possibility, i.e., INOV positively correlating with $\beta_{app}$ has been hinted in some INOV studies \citep{2005MNRAS.356..607S, 2019MNRAS.483.3036O}, the present study demonstrates this link, for the first time with statistical robustness, based on an extensive INOV campaign focused on a blazar sample selected specifically for addressing this question (see Section~\ref{sec:introduction}). Here, it may be reiterated that our TeV-HBLs do exhibit the other common trait of blazars, namely a substantial fractional polarization ($p_{opt}$ > 3\%, Table~\ref{tab:sample_table}; Section~\ref{sec:introduction}). Not only is the maximum recorded $p_{opt}$ consistent with this lower limit, but so is the mean value $p_{opt}$ (except in the case of the blazar J0136+3905, but here too, $p_{opt}$ was found to be above 3\% in 2 out of the total 7 measurements available in the RoboPol survey (\citealt{2021MNRAS.501.3715B})
%Even the single possible exception to INOV non-detection, J050756+673724, recorded $p_{opt}$ above the 3\% threshold in 2 out of the 7 available measurements (maximum $p_{opt}$  = 6.5+- 0.4\%, Table A1) {\color{red} (HOW?, above the 3\% threshold in 11 out of 15 measurements in ROBOPOL)} and therefore this blazar too is not incongruous with an HPQ classification
{\footnote{Note that, unless the number of measurements is very large, the maximum value of $p_{opt}$ may be preferred over the mean value, as this would reduce the chance of missing out genuine blazars/HPQs since their polarization is known to vary and hence might average below the defining threshold of 3\% due to frequent dips \citep{1991ApJ...375...46I,2022MNRAS.516L..18C}.}}).
In this context, we further note that the high-polarization TeV-HBL J0416+0105 of our sample, having p$_{opt}$ (mean) = 6.3$\pm$0.3\% did not exhibit INOV (down to the 2\% detection limit), not only in the 4 sessions reported here, but also in the 8 sessions (2016-18) reported by \citet{2020MNRAS.496.1430P}.
\par
%%%%%%%%%%%%%%%%%%%%%%%%%%%%%%%%%%
It is also noteworthy that the non-detection of INOV even at $\sim$2\% level for essentially our entire sample of TeV-HBLs, also circumscribes the role of accretion disk flares (or instabilities) as a possible cause of INOV \citep[e.g.,][]{1993ApJ...406..420M, 1993ApJ...411..602C}, at least in the case of geometrically thick  radiatively-inefficient disks that are supposed to fuel intrinsically low-power AGN like the TeV-HBLs being discussed here (Section~\ref{sec:introduction}). \par
%%%%%%%%%%%%%%%%%%%%%%%%%%%%%%%%%%
As mentioned in Section~\ref{sec:introduction}, the two main factors perceived to be responsible for INOV of jet-dominated AGN (blazars) are: (i) injection/growth of turbulence within the jet plasma whose synchrotron emissivity and fractional polarization get enhanced while passing through one or more shocks (e.g., \citealt{2014ApJ...780...87M} and references therein; \citealt{2015JApA...36..255C, 2016ApJ...820...12P, 2021Galax...9..114W}; see also, \citealt{2012A&A...544A..37G,1980MNRAS.193..439L}); and (ii) the bulk Doppler factor $\delta_{j}$ of the post-shock turbulent plasma in the jet. While the former physical process is crucial for the origin of micro-variability (INOV) of the jet's emission, the latter factor holds the key to the INOV detection (via a Doppler boost). 
Clearly, it is important to find observational basis for this scenario. A decade ago, \citet{2012A&A...544A..37G} investigated the dependence of INOV on fractional optical polarization ($p_{opt}$), by carrying out sensitive, high-cadence optical monitoring of 21 radio-loud quasars, including 9 high- and 12 low-polarization quasars (HPQs 
and LPQs), taking the dividing line at the conventional $p_{opt}$ = 3\% \citep[e.g.,][]{1984ApJ...279..485S}. Remarkably, the HPQ subset showed strong INOV (amplitude 
$\psi$ > 4\%) on 11 out of 29 nights, in stark contrast to the LPQs for which strong INOV was observed on just 1 out of 44 nights. This clearly established a high $p_{opt}$ as a key attribute of the radio quasars showing strong INOV. But, is this alone a sufficient marker for detection of strong INOV? What about the role of the afore-mentioned second factor, namely, $\delta_{j}$ ? Indeed, an observational hint for such a correlation was noticed in a recent INOV study of 3 narrow-line Seyfert1 galaxies \citep{2019MNRAS.483.3036O}. The results presented here place such a correlation on a statistically firm footing, for the first time, by focusing on an extreme subset of blazars, namely TeV-HBLs, whose nuclear radio jets exhibit only slow moving (at most mildly superluminal) features, consistent with small Doppler boosting of their emission. For this subset, the present work demonstrates an essentially total lack of INOV detection.
%The near-absence of INOV detection for TeV-HBLs  
This can be readily understood if the apparent kinematics of these dominant radio knots, which appear at most mildly superluminal, reflects the bulk motion of the underlying jet (at least the sheath layer), as argued by \citet{2009AJ....138.1874L, 2009ApJ...696L..22L} and others \citep[e.g.,][]{2009ApJ...696L..17K, 2010ApJ...722..197L}. In this framework, compared to the highly superluminal radio knots typically observed in blazar jets, the dominant slow moving radio knots observed in the VLBI jets of TeV-HBLs would have to be much more luminous intrinsically, in order to be detectable even without the benefit of a strong Doppler boosting. On the other hand, this would not be required in case the VLBI knots are mere `patterns’, kinematically decoupled from the underlying (much faster) jet, as suggested in several studies (e.g., \citealt{2018ApJ...853...68P} and references therein;
\citealt{1997ARA&A..35..607Z,2004ApJ...609..539K}).
%2007ApJ...668L..27K, 2008ApJ...680..867K, 2009ApJ...699.1274B, 2014ApJ...787..151C, 2016A&A...592A..22H% {\color{red}(TO CHECK ALL)}. 
In that event, the observed brightness of the radio knots and the level of INOV originating in the jet’s turbulent zone, would both be dictated by the beaming associated with the bulk velocity of the underlying jet (the, so called,  `emission velocity’ of the jet, cf. \citealt{1979ApJ...232...34B}), despite little direct evidence for a relativistic flow coming from VLBI observations. The rather tight correlation of INOV with the apparent speed of the VLBI knots, as found here, suggests that at least the zone of turbulence within the (post-shock) jet-flow remains kinematically coupled to the (slow moving) shock/knot, perhaps due to entanglement of the magnetic field lines, regardless of whether the observed kinematics of such shocks reflects the bulk speed of the underlying jet. \par
%%%%%%%%%%%%%%%%%%%%%%%%%%%%%%%%%%%%%

Finally, it should be emphasized that the very low INOV duty cycle inferred here for TeV-HBLs represents a `population characteristics’ and it is not meant to be a permanent metric for the INOV of any individual member of this class of blazars, e.g., by implying that no such blazar would ever exhibit a strong INOV. This important point has been underscored in \citet{1999A&AS..135..477R} by highlighting the case of the prominent TeV-HBL PKS 2155-304.
% which has been a part of multiple INOV campaigns. 
This blazar, well-known for ultra-rapid variability of its TeV emission, is prone to slipping into prolonged spells of INOV quiescence, as noted by these authors. 
%Whether such INOV state transition is also reflected in the apparent motion of the VLBI radio knot(s), would be an interesting observational aspect to probe in future VLBI campaigns. 
Another such example, the TeV-HBL PG1553+111, is a member of the present sample itself (Table~\ref{tab:sample_table}). Its low INOV duty cycle implied by the non-detection of INOV on all 4 nights during 2022 (online Fig.~S6) is statistically compatible with its recent study by \citet{2023MNRAS.519.2796D} in which INOV was detected on just 4 out of 27 nights of R-band monitoring during 2019. In contrast, during 2009--10, this blazar exhibited strong INOV ($\psi \gtrsim$ 5\%) on all 3 nights it was monitored in R-band \citep{2011MNRAS.416..101G}. This indicates a transition to INOV quiescence, occurring somewhere between 2009-10 and 2019-22. Although a detailed comparison of this pattern with the jet's kinematic on parsec scale is currently lacking, it is interesting to note that the published MOJAVE images at 15 GHz do indicate a drop in the apparent speed of the dominant VLBI knots by a factor of $\sim 3$ over the period from 2008 to 2018 \citep[see Fig. 1 of][]{2017ApJ...851L..39C}, which is consistent with the above-inferred change in the INOV state (from high to low) of this TeV blazar. It would be desirable to garner further evidence on the question whether INOV state transitions are accompanied by a changing kinematics of the parsec-scale radio jets.
%%%%%%%%%%%%%%%%%%%%%%%%%%%%%
%%%%%%%%%%%%%%%%%%%%%%%%%%%%%
%%%%%%%%%%%%%%%%%%%%%%%%%%%%%%%%%%%%%
\section{CONCLUSIONS}
\label{sec:conclusions}
We have carried out an extensive, high-sensitivity intranight optical monitoring programme targeted on a well-defined sample of 6 TeV detected HBLs whose parsec-scale jets 
had been shown to be dominated by radio knots exhibiting either subluminal, or at most mildly superluminal motion. An essentially zero INOV duty cycle is estimated here from the 24 monitoring sessions devoted to these TeV-HBLs, despite their exhibiting fairly high degree of optical polarization. This INOV duty cycle is at least an order-of-magnitude lower than that typical of radio-selected blazars (LBLs, whose parsec-scale jets are usually dotted with highly superluminal knots, e.g., \citealt{2007ApJ...658..232C, 2007A&A...472..763B, 2009AJ....138.1874L,2017ApJ...846...98J}). Thus, TeV-HBLs with slow-moving VLBI knots are clearly identified for the first time as an extreme sub-population of blazars, from the perspective of INOV. Their highly subdued INOV, as found here, demonstrates that the presence of dominant superluminal radio knot(s) in the parsec-scale jet constitutes a key diagnostic for INOV detection and while a high degree of optical polarization is also an important marker, as shown in \citet{2012A&A...544A..37G}, it alone is not a sufficient diagnostic for INOV detection.
%%%%%%%%%%%%%%%%%%%%%%%%%%%%%%%%%%%%%
%%%%%%%%%%%%%%%%%%%%%%%%%%%%%%%%%%%%%
%%%%%%%%%%%%%%%%%%%%%%%%%%%%%%%%%%%%%
%%%%%%%%%%%%%%%%%%%%%%%%%%%%%%%%%%%%%
%%%%%%%%%%%%%%%%%%%%%%%%%%%%%
%%%%%%%%%%%%%%%%%%%%%%%%%%%%%
\section*{Acknowledgements}

We thank the anonymous referee for the valuable comments on our manuscript. GK would like to thank Indian National Science Academy for a Senior Scientist position. The assistance from the scientific and technical staff of ARIES DFOT is thankfully acknowledged. VN thanks Krishan Chand, Nikita Rawat, Bhavya Ailawadhi and Sriniwas M Rao for help with observations.
%%%%%%%%%%%%%%%%%%%%%%%%%%%%%%%%%%%%%%%%%%%%%%%%%%%%%%%%%%%%%%%%%%%%%%%%%%%
\section*{Data Availability}
The data used in this study will be shared on reasonable request to the corresponding author.
\bibliographystyle{mnras}
\bibliography{references} 
\bsp	% typesetting comment
\label{lastpage}
\end{document}